\documentclass{PoS}

\usepackage{amssymb}
\usepackage{amsmath}

\usepackage{graphicx}

\newcommand{\be}{\begin{equation}}
\newcommand{\ee}{\end{equation}}
\newcommand{\bea}{\begin{eqnarray}}
\newcommand{\eea}{\end{eqnarray}}
\newcommand{\no}{\noindent}
\newcommand{\nn}{\nonumber}

\newcommand{\Tr}{{\rm Tr\,}}
\newcommand{\Det}{{\rm Det\,}}
\newcommand{\e}{{\rm e\,}}

\newcommand{\rar}{\rightarrow}

\title{The High Density Region of QCD from an Effective Model }

\ShortTitle{The High Density Region of QCD from an Effective Model}

\author{Roberto De Pietri\\
        Dipartimento di Fisica, Universit\`a di Parma and
        INFN Gruppo Collegato di Parma, Italy \\
        E-mail: \email{roberto.depietri@fis.unipr.it}}

\author{\speaker{Alessandra Feo}\thanks{Work partially supported by 
        INFN Gruppo Collegato di Parma.}\\
        Dipartimento di Fisica Teorica, Universit\'a di Torino and 
        INFN Sezione di Torino, Italy  \\
        E-mail: \email{feo@to.infn.it}}
 
\author{Erhard Seiler\\
        Max-Planck-Institut f\"ur Physik (Werner Heisenberg Institut), 
        M\"unchen, Germany \\
        E-mail: \email{ehs@mppmu.mpg.de}}

\author{Ion-Olimpiu Stamatescu\\
        FEST - Protestant Institute for Interdisciplinary Research, 
        Heidelberg and 
        Institut f\"ur Theoretische Physik  der Universit\"at, Heidelberg, 
        Germany \\
        E-mail: \email{I.O.Stamatescu@thphys.uni-heidelberg.de}}

\abstract{We study the high density region of QCD within an effective model 
obtained in the frame of the hopping parameter expansion and choosing 
Polyakov-type loops as the main dynamical variables representing
the fermionic matter. This model still shows the
so-called sign problem, a difficulty peculiar to non-zero chemical
potential, but it permits the development of algorithms which ensure a
good overlap of the simulated Monte Carlo ensemble with the true one. 

We review the
main features of the model
and present results concerning the dependence of
various observables on the chemical potential and on the temperature, in
particular of the charge density and the Polykov loop susceptibility, which may
be used to characterize the various phases expected at high baryonic
density. In this way, we obtain information about the phase structure of the
model and the corresponding phase transitions and cross over regions, which
can be considered as hints about the behaviour of non-zero density QCD.

\begin{flushright}
DFTT 24/2007
\end{flushright}
}

\FullConference{The XXV International Symposium on Lattice Field Theory\\
                 July 30-4 August 2007\\
                 Regensburg, Germany}

\begin{document}

\section{Introduction}
The exploration of the phase diagram of QCD at non-zero baryon density is a challenging 
and interesting problem. In particular, it has been emphasized that quark 
matter at extreme high density may behave as a color superconductor 
and it is also expected that the phase diagram in the 
temperature-density plane shows multiple phases separated 
by various transition lines but very little is known about their exact
position and nature. Lattice calculations, using different approaches that try to
evade the sign problem generated by the non-zero chemical potential, have been
mostly implemented at high temperature and small baryon density, where they agree
reasonably well with each other. In this region there is good evidence for the presence 
of a crossover instead of a sharp deconfining transition.
At large $\mu$ (baryon density), however, there are only few numerical results 
which need to be corroborated by using different methods.
See \cite{karrev} for a review.

The purpose of this work is to get further insight about the phase structure of high
density, strongly interacting matter using an effective model. 
In the spirit of the $\mu=0$ quenched approximation a `non-zero density
quenched approximation' for $\mu > 0$ based on the double limit 
$ M \rar \infty,\, \mu \rar \infty,\, \zeta \equiv {\rm exp}\,(\mu - \ln
M) :$ fixed \cite{bend,fktre} has been considered. This implements a 
static, charged background, which influences the gluonic dynamics
\cite{fktre,bky}. The present model \cite{hdm01} represents a
systematic extension of the above considerations: the gluonic vacuum
is enriched by the effects of dynamical quarks of large (but not
infinite) mass, providing a large net baryonic charge.  
In \cite{hs,dfss2,dfss} we explore the phase structure of the model,
as a first step in understanding the properties of such a background.

\section{QCD at Large Chemical Potential}

We use the QCD grand canonical partition function with Wilson fermions 
at $\mu>0$ defined as:
\begin{align}
& \!\!\!\!\!\!\!\!\!\!\!\!\!\!\!
{\cal Z}(\beta,\kappa,\gamma_G,\gamma_F,\mu) = \int[DU]\, 
\e^{-S_G(\beta,\gamma_G,\{U\})}{\cal Z}_F({ {\kappa}},\gamma_F, \mu, \{U\}) \, ,
\\ 
& \!\!\!\!\!\!\!\!\!\!\!\!\!\!\!
S_G(\beta,\gamma_G,\{U\})  
= -\frac{\beta}{N_c}\,Re\,\Tr\,\left(\frac{1}{\gamma_G}\,
\sum_{j>i=1}^3\,
P_{ij} + \gamma_G\,\sum_{i}\,P_{i4}\right)\, , \\ 
& \!\!\!\!\!\!\!\!\!\!\!\!\!\!\!
{\cal Z}(\beta,\kappa,\mu) = \int[DU]\, 
\e^{-S_G(\beta,\{U\})}{\cal Z}_F({ {\kappa}}, \mu, \{U\}) \, ,
{\cal Z}_F({ {\kappa}}, \mu, \{U\}) =  
\Det W ({ {\kappa}}, \mu, \{U\}) \, , \\ 
\begin{split} 
W_{ff'} &= \delta_{ff'} [ 1 -  \kappa_f\, \sum_{i=1}^3 \left( 
\Gamma_{+i}\,U_i\,T_i +
\Gamma_{-i}\,T^*_i\,U^*_i\right) 
- \kappa_f\,  \left( \e^{\mu_f}\,\Gamma_{+4}\,U_4\,T_4 +
\e^{-\mu_f}\,\Gamma_{-4}\,T^*_4\,U^*_4 \right) ] \, ,  \\
\Gamma_{\pm \mu} &= 1 \pm \gamma_{\mu},\ \ \gamma_{\mu}=\gamma_{\mu}^*,\
\gamma_{\mu}^2=1 \,\, ,
\kappa = \frac{1}{2(M+3+\cosh \mu)} = \frac{1}{2(M_0+4)} \, ,
\end{split} \nn
\end{align}
\no where we used $S_G$ for Wilson's plaquette $(P)$ action and 
used a certain definition of the Wilson term in $W$.
Here $M$ is the `bare mass', $M_0$  the bare mass at $\mu=0$,
$f$ is the flavor index, $U_\mu$ the link variables and $T_\mu$ 
lattice translations. A non-zero physical temperature $T$ is 
introduced as $a T = \frac{\gamma_{phys}}{N_\tau}$, where 
$\gamma_{phys}$ is the physical cutoff anisotropy defined by an
appropriate renormalization of the coupling anisotropies,
and $N_\tau$ the `length' of the (periodic) temporal lattice size.

\begin{figure}
\begin{center}
\includegraphics[width=.6\columnwidth]{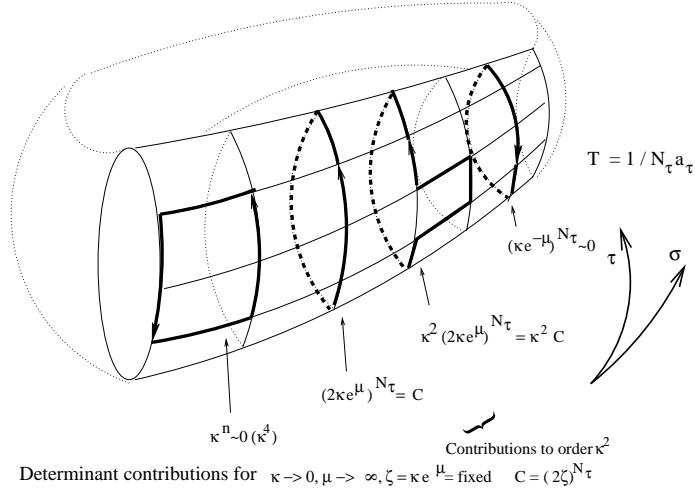}  
\end{center}
\caption{Periodic lattice, loops, temporal gauge. In the maximal temporal gauge
also the links of the basis line are fixed to 1 up to the rightmost one.} 
\label{torus_2} 
\end{figure}

At large $\mu$ the behavior of QCD quantities may however be dominated by 
certain factors in the fermionic determinant which lead to a simpler model 
that is actually easier to simulate.
The model we study is based on an analytic expansion of QCD (the 
hopping parameter expansion) up to second order in $\kappa$ 
(see \cite{dfss} for details) 
and its main ingredient are Polyakov-type 
loops (see fig. \ref{torus_2}),
capturing the effect of heavy quarks with low mobility.
The model still has a sign problem, being the fermionic determinant
complex at $\mu \ne 0$, but being based on the variables which are 
especially sensitive to the physics of dense baryonic matter it 
allows for reweighting algorithms which ensure a good overlap 
of the Monte Carlo ensemble with the true one. 

We use the Wilson action and Wilson fermions within a reweighting 
procedure. The updating is performed with a local Boltzmann factor which 
only leads to a redefinition of the ``rest plaquette'':
\be
B_0(\{U\})\equiv\prod_{Plaq} e^{\frac{\beta}{3} Re \Tr Plaq} 
\prod_{\vec x}  \exp\bigg\{ 
2 C Re \Tr \bigg[ {\cal P}_{\vec x} + \kappa^2 \sum_{i,t,t'}
{\cal P}^{0,1}_{\vec x,i,t,t'} \bigg] \bigg\} \, .
\ee
The weight (global, vectorizable) is
\be
w(\{U\})\equiv \prod_{\vec x} \exp\bigg\{ 
\!\!  -2 \, C\, Re \Tr \bigg[ {\cal P}_{\vec x} + \kappa^2 
\sum_{i,t,t'}
P^{0,1}_{\vec x,i,t,t'} \bigg] \bigg\} {\cal Z}^{[2]}_F(\{ U \}) \, ,
\ee
such that the 'Boltzmann factor' becomes,
\be
w\,B_0\, = \, B\,\equiv \prod_{Plaq} e^{\frac{\beta}{3}Re \Tr 
Plaq}\,{\cal Z}^{[2]}_F(\{ U \}) \, .
\ee
Averages are calculated by reweighting according to,  
$B = B_0 w_0$ 
and 
$ \langle O \rangle = \frac{\langle w_0 O\rangle_0}{\langle w_0 \rangle_0}$. 

We have employed the Cabibbo-Marinari 
heat-bath procedure mixed with over-relaxation. This updating already 
takes into account part of the $\mu > 0$ effects and the generated 
ensemble can thus have a better overlap with the true one than an updating 
at $\mu = 0$. One can also use an improved $B_0$, to be taken care of by a 
supplementary Metropolis check. 
Notice that extracting a factor like $B_0$ may also improve 
convergence of full QCD simulations at $\mu > 0$.

We measure several observables under the variation of $\mu$ and 
$T$, to check the properties of the different phases for small $T$ and large
$\mu$. In the following we specialize to $N_c=3$. The observables are: 
the Polyakov loop ($\langle P \rangle$) and its susceptibility ($\chi_P$)
\bea
\langle P \rangle =  \langle \frac{1}{3\,N_\sigma^3} \sum_{\vec x} \Tr {\cal 
P}_{\vec x}\rangle = \langle \frac{1}{N_\sigma^3} \sum_{\vec x} P_{\vec x}\rangle  
\,,
\qquad
\chi_P=\sum_{\vec y} \left( \langle P_{\vec x}\, P_{\vec y}\rangle
- \langle P_{\vec x}\rangle \langle P_{\vec y}\rangle\right)\, ,
\eea
the (dimensionless) baryonic number density 
$n_B= \sum_f \frac{n_{b,f}}{T^3}$,
with the corresponding 
susceptibility. To check the character of 
the conjectured third phase we also measure (but we do not report the
corresponding results here) the spatial and temporal 
plaquettes, the topological charge, topological charge susceptibility 
and the diquark-diquark correlators (see \cite{dfss} for details).

The simulations are mainly done on lattice $6^4$ for $n_f=1,3$ degenerate 
flavors (any mixture of flavors can be implemented). The $\kappa$ 
dependence has been analyzed in \cite{hdm01}. Here we set $\kappa=0.12$ 
(rather ``small'' bare mass $M_0 = 0.167$) which drives the $1/M^2$ 
effects in the baryonic density to about $50 \%$. 
The task we have set to 
ourselves is primarily to explore the phase structure of the model at 
large chemical potential and ``small'' temperature and we
accordingly vary $\mu$ and $\beta$.


\section{Results and Discussions}

\begin{figure}
\includegraphics[width=.33\columnwidth]{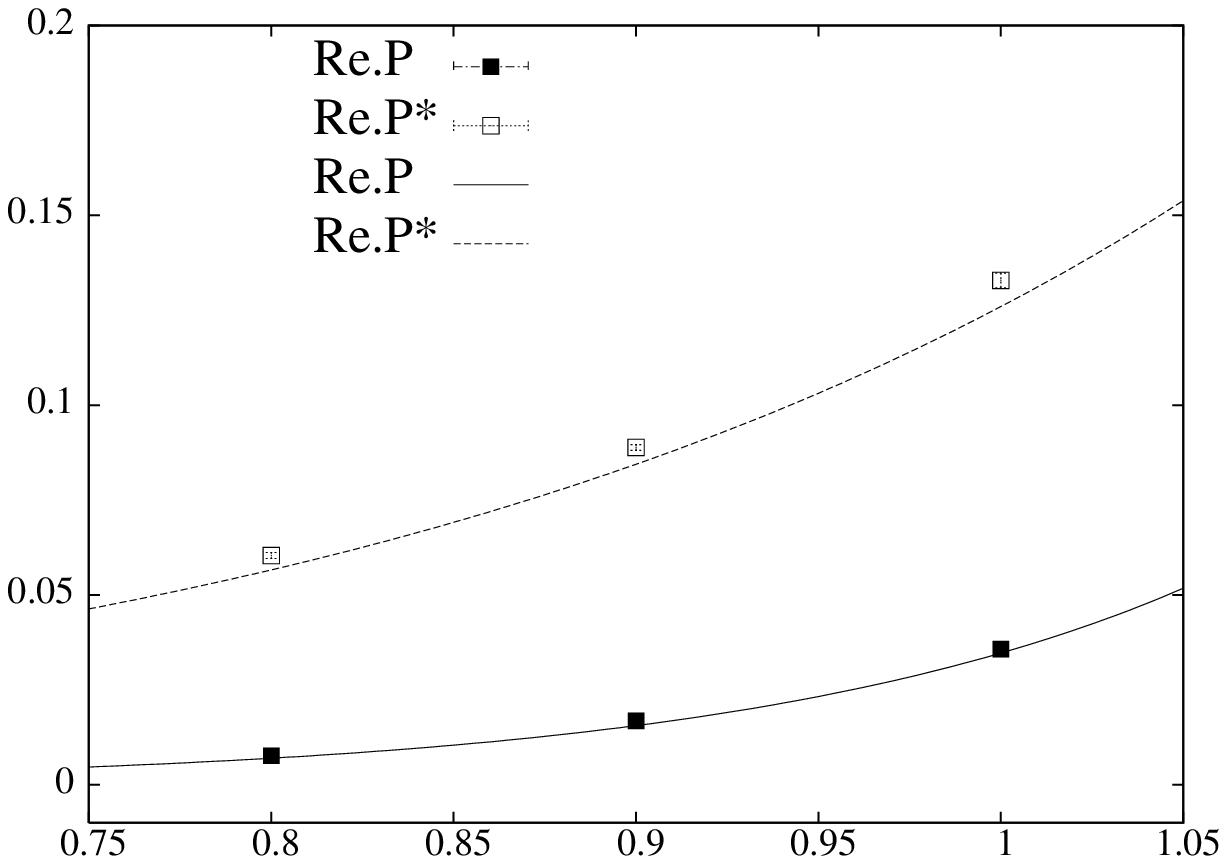}
\includegraphics[width=.33\columnwidth]{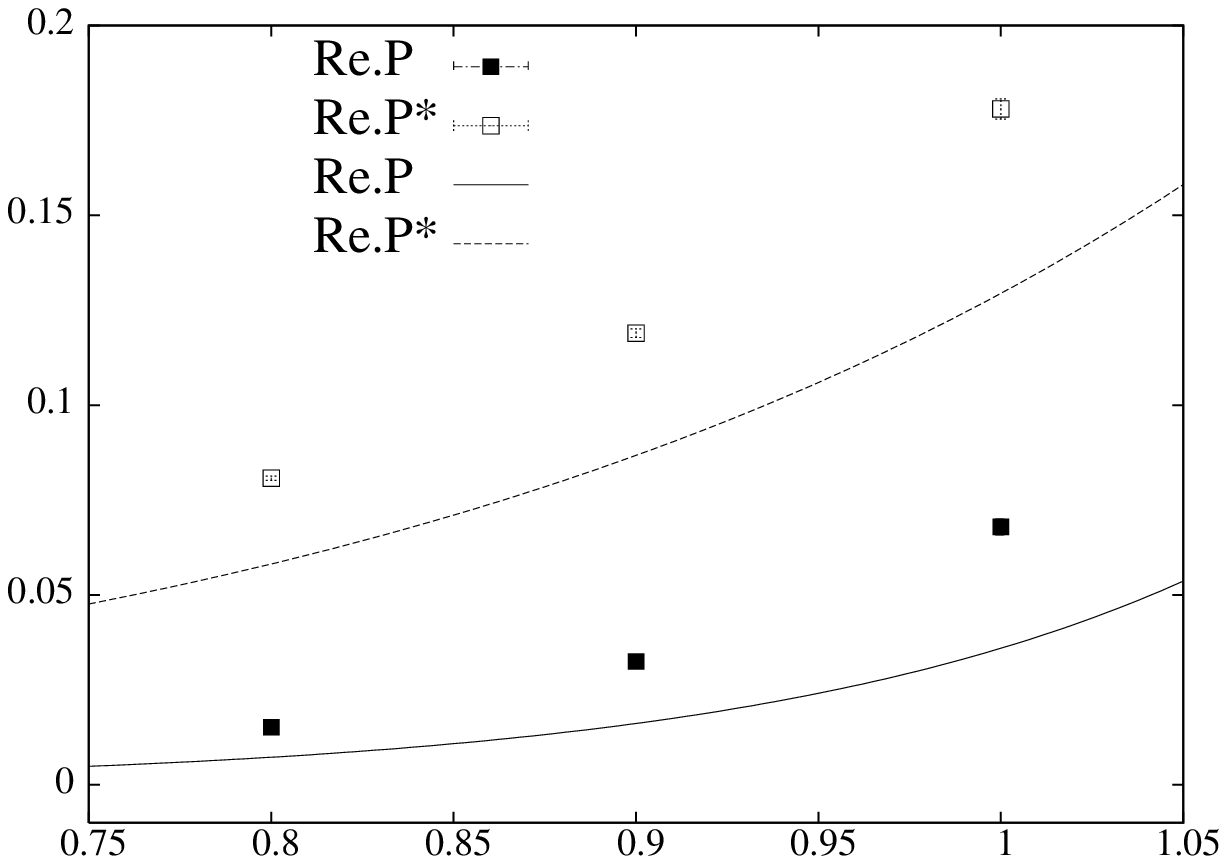}
\includegraphics[width=.33\columnwidth]{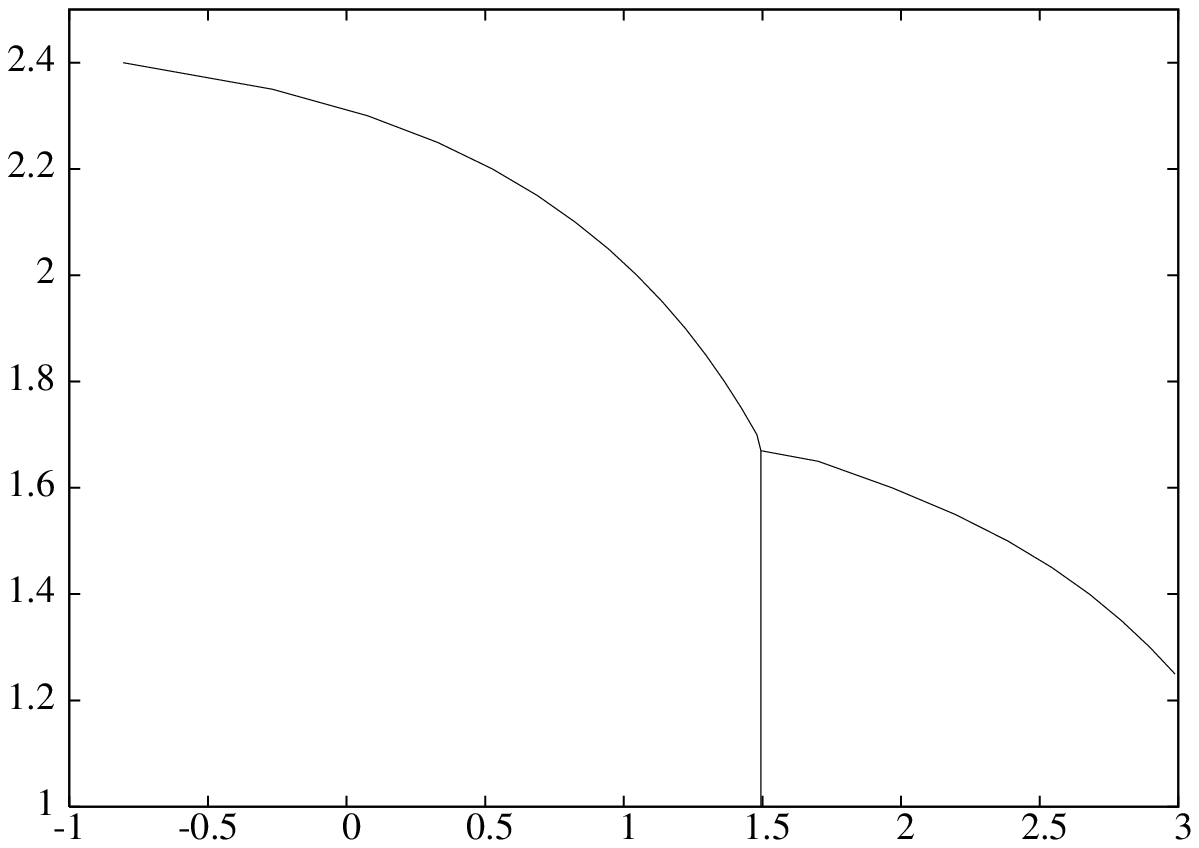}
\caption{Comparison with strong coupling at $\beta=3$ (left plot) and
$\beta=5$ (center plot), $4^4$ lattice. Full symbols denote $Re P$, 
empty symbols $Re P^\ast$, the lines show the corresponding strong 
coupling results. On the right is reported the mean field phase diagram 
(abscissa $\mu$, ordinate $\gamma=N_\tau\, a\,T$).}
\label{sc443}
\end{figure}

As a first orientation about the behavior of the model we consider the 
strong coupling/hopping parameter expansion, which also serve as a check
of the Monte Carlo results. 
In fig. \ref{sc443} we compare the Monte Carlo results of the 
Polyakov loop and its adjoint 
on $4^4$ and $6^4$ lattices, for $\kappa=0.12$, one flavor and 
different values of $\beta$, with $P^{[2]}$ and $P^{\ast [2]}$
(where $[2]$ means second order in the strong coupling expansion,
see \cite{dfss} for details). 
The agreement is good for the $4^4$ lattice and $\beta=3$, 
while for $\beta=5$ there are already significant deviations
which show strong effects at large $\mu$ even at moderate $\beta$ that 
may indicate possible phase transitions. 
But the agreement between Monte Carlo and strong coupling 
results is sufficient to validate the simulations. 

We also performed mean field calculations using a temporal gauge 
fixing \cite{dfss} and introducing two different mean field 
variables for the spatial component ($u$) and the temporal component
($v$) of the gauge field. 
They give some qualitative insight into the phase
structure of the model to which Monte Carlo simulations can be 
compared. In fig. \ref{sc443} we give an illustrative example, taken with 
$\beta=4$ and $N_\tau=6$. It shows a large `confinement' region for small 
$T$ and $\mu$ corresponding to the trivial fixed point mentioned above 
with both mean fields $u$ and $v$ vanishing. For larger $T$ or $\mu$ one 
crosses into a deconfined regime with both mean fields $u,v> 0$. In the 
lower right corner there appears in addition an intermediate phase with 
$u=0,\ v>0$. The field $v$ is close to its maximal value 1 wherever it 
is nonzero, whereas $u$ has smaller, varying values, depending on the 
region.

\begin{figure}
\begin{center}
\includegraphics[width=0.42\columnwidth]{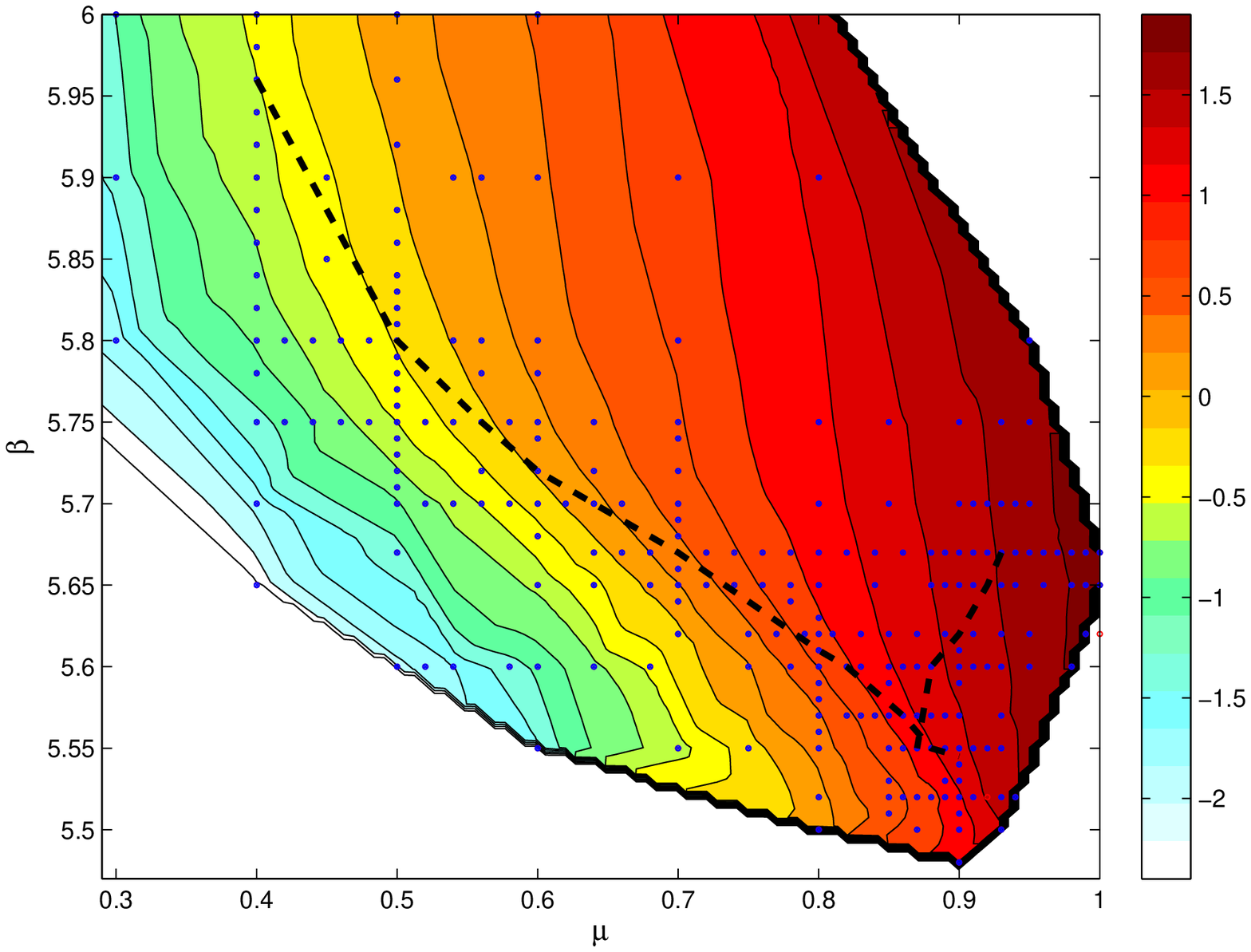}
\hspace{5mm}
\includegraphics[width=0.42\columnwidth]{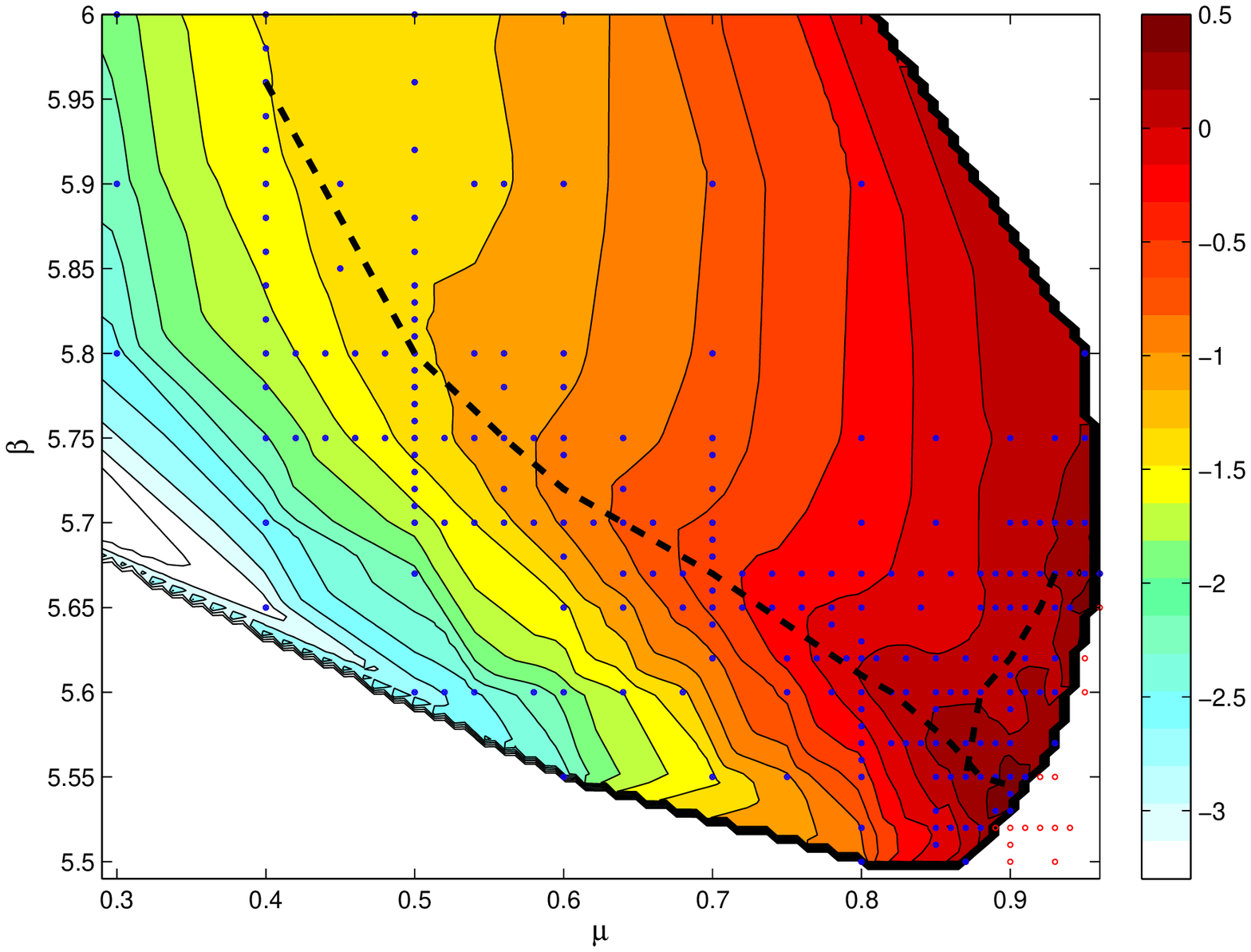}
\end{center}
\vspace{-0.5cm}
\caption{Left: Landscape of the baryonic density. Right:
Landscape of the baryon density susceptibility.
The color scale (right) is based on $\log_{10} (n_B)$.}
\label{densR}
\end{figure}

\begin{figure}
\begin{center}
\includegraphics[width=.995\columnwidth]{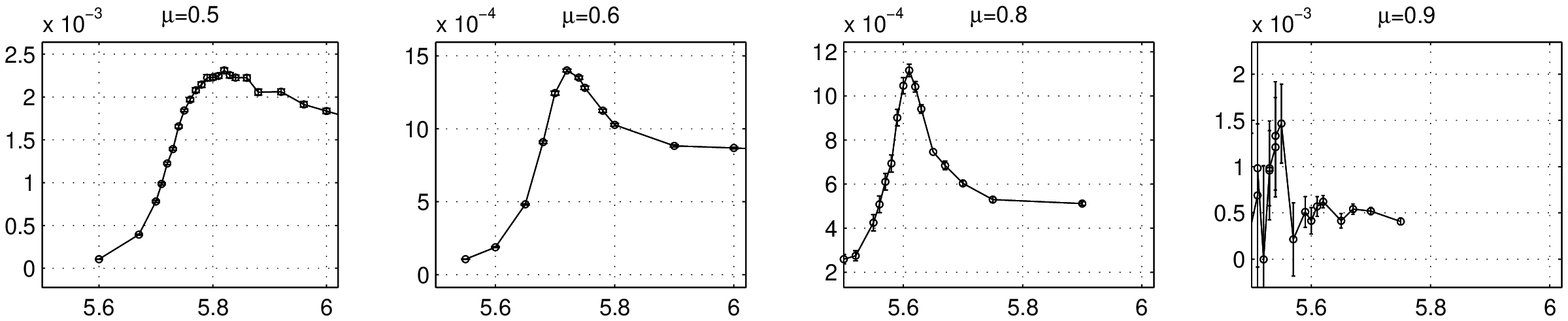}
\includegraphics[width=.995\columnwidth]{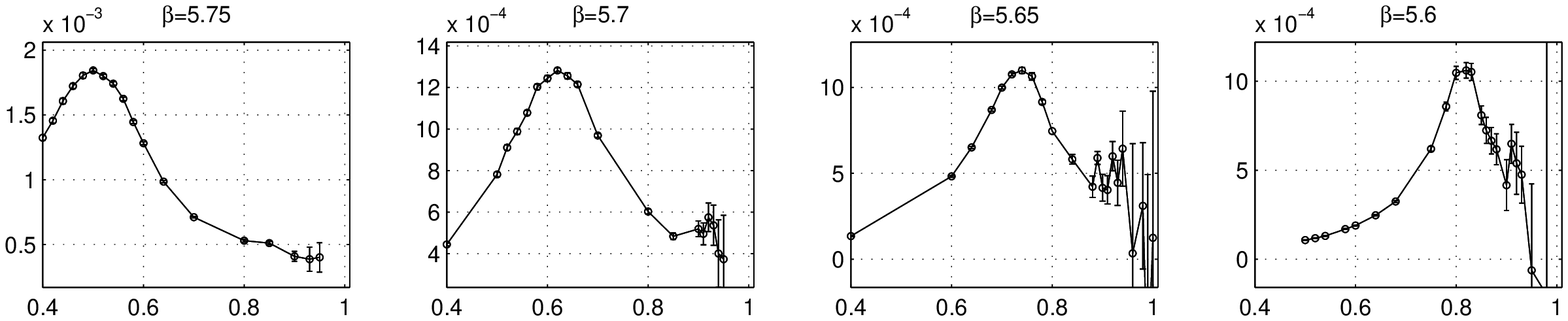}
\end{center}
\vspace{-0.5cm}
\caption{Polyakov loop susceptibility vs. $\beta$ at fixed $\mu$ (top)
and vs. $\mu$ at fixed $\beta$ (bottom). }
\label{poleps}
\end{figure}

\begin{figure}
\includegraphics[width=0.55\columnwidth]{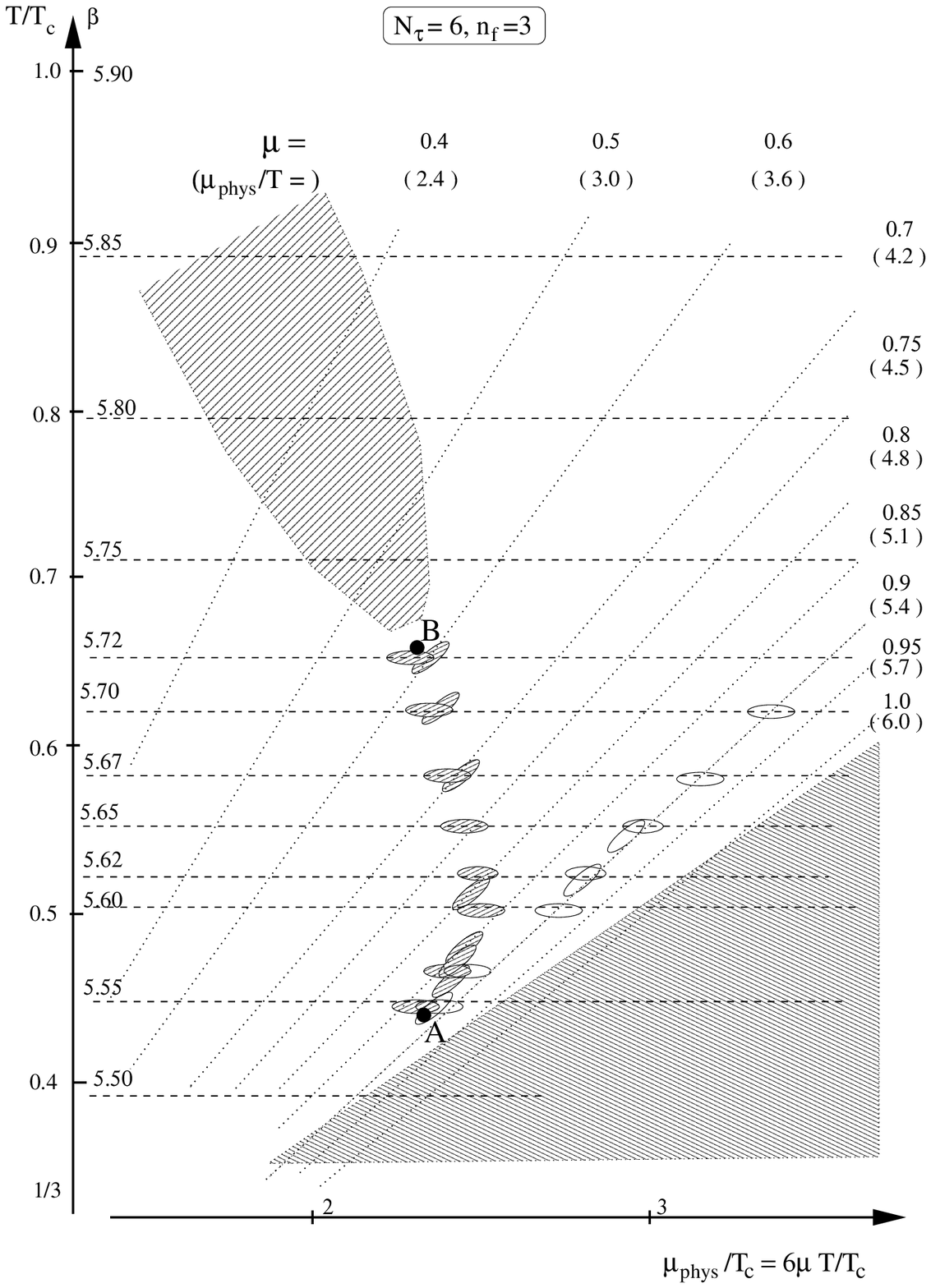}
\hspace{1mm}
\hbox{\vbox{
\hbox{\includegraphics[width=0.42\columnwidth]{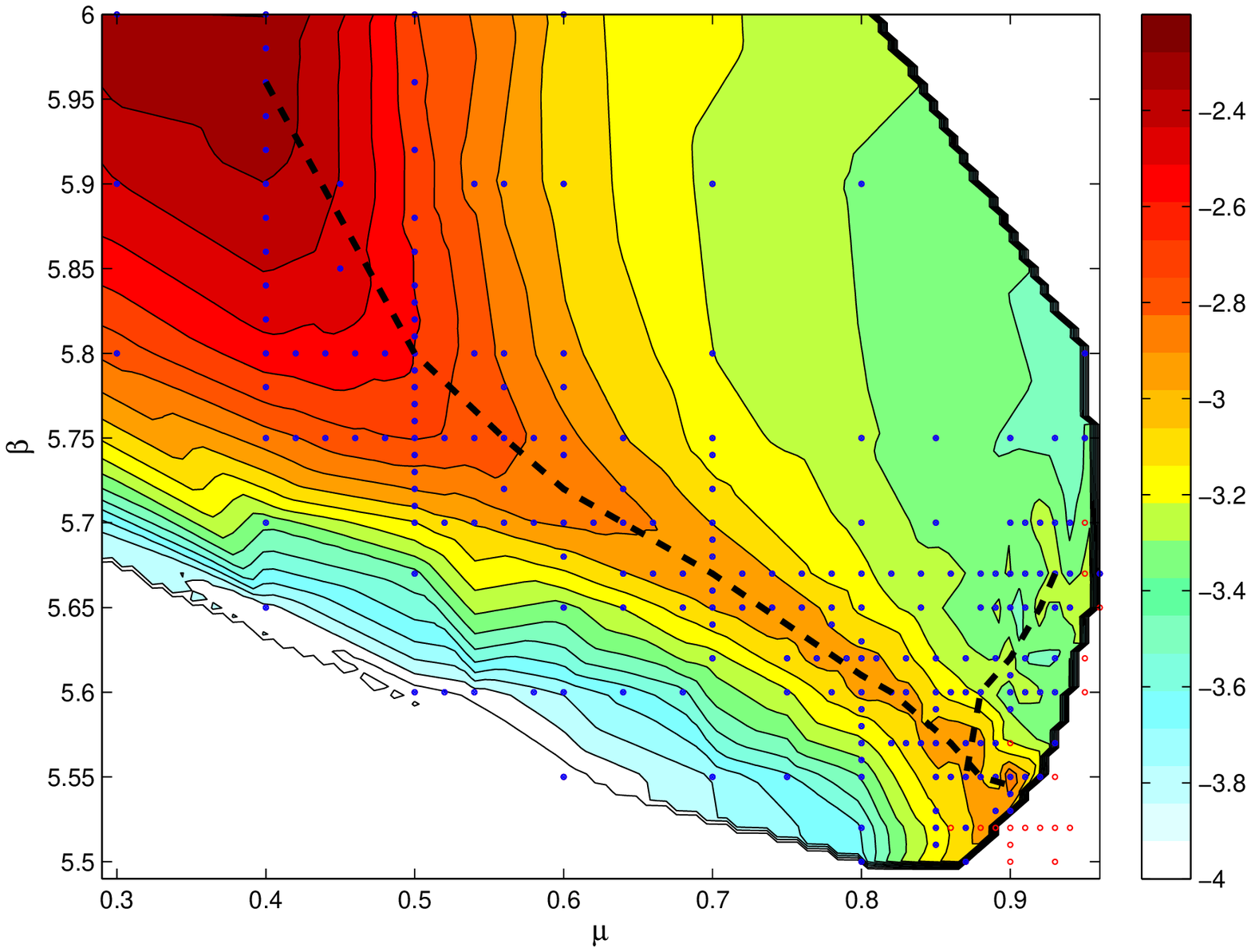}}
\vspace{.8cm}
\hbox{\includegraphics[width=0.42\columnwidth]{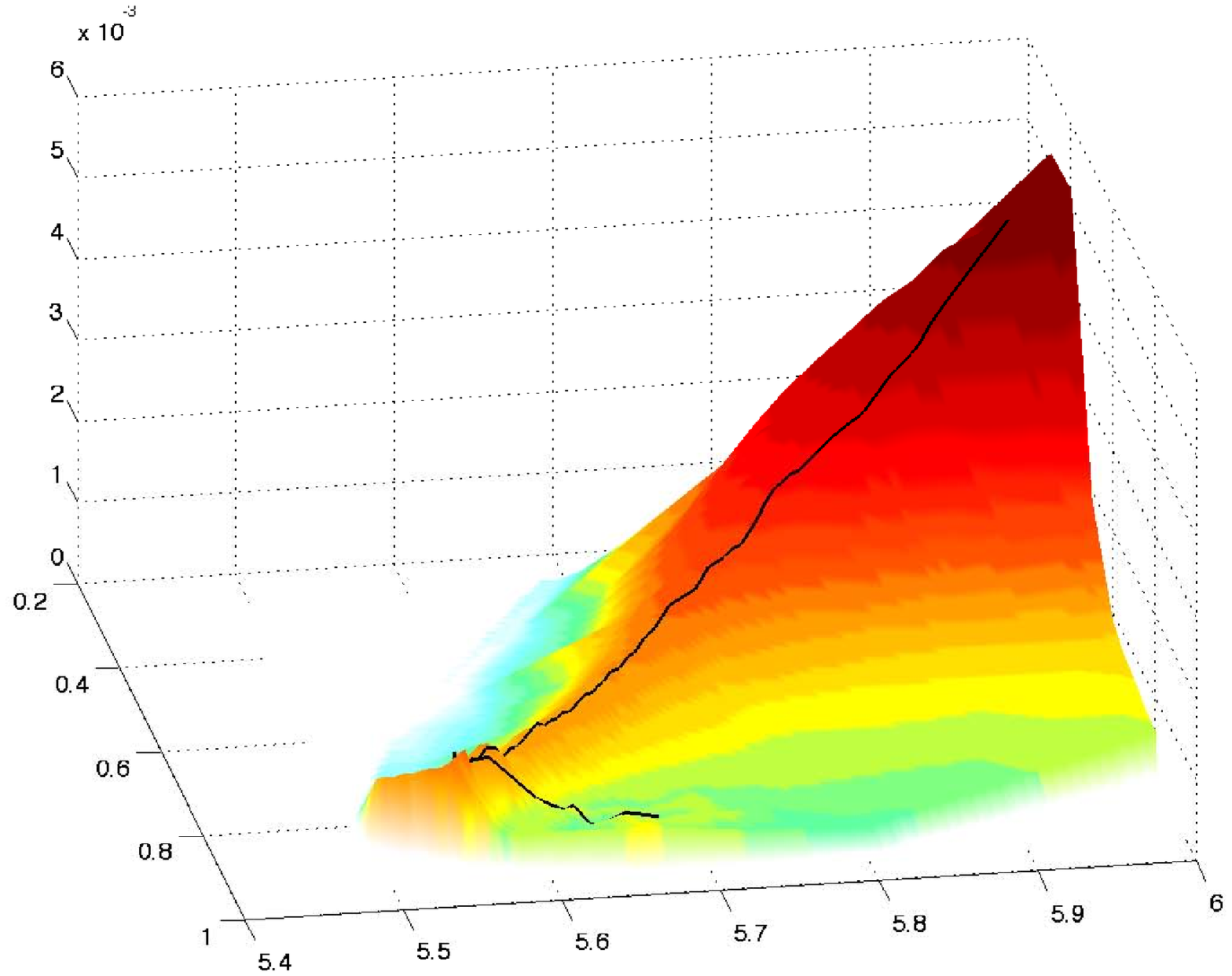}}
\vspace{1cm}
}}
\vspace{-0.2cm}
\caption{Left: the phase diagram in the  $\beta$ (or $T/T_c$) - $\mu_{phys}/T_c$ QCD
plane. The dotted straight lines correspond to constant $\mu$, 
the dashed ones to constant $\beta$. Right: landscape of the Polyakov loop 
susceptibility (top) and its 3d view (bottom). The color scale
is based on $\log_{10} (\chi_P)$.} 
\label{polland}
\end{figure}

The algorithm works reasonably well over a large range of parameters 
even at small temperature. The model permits to vary $\mu$, $\kappa$,
$\beta$ as independent parameters and it is reasonably cheap to measure
various correlations. 
In fig. \ref{densR} we show the ``landscape'' of the real part of the 
baryon density $n_B$ (while the imaginary part is compatible with zero
inside the statistical errors). The main variation is an exponential 
growth with $\mu$ indicating that we do not see yet saturation effects.
This masks to a certain extent the finer structure. 
A clearer view of the situation is provided by looking at the 
``landscape'' of the susceptibility of the baryon density 
(fig \ref{densR}). A ridge is clearly visible, highlighted by a 
dashed black line. A second line (dashed) is explained later. 
We found it 
therefore advantageous to look at the Polyakov loops and their 
susceptibility. 
In fig. \ref{poleps} we show the Polyakov loop susceptibility
vs. $\beta$ at fixed $\mu$ and on the bottom vs. $\mu$ 
at fixed $\beta$ and,
in fig \ref{polland} (right) the corresponding landscape. 
The plots of the Polyakov loop susceptibility show quite clearly maxima
indicating possible transitions or crossovers. In the landscape 
fig. \ref{polland}, one of these maxima shows up as a well defined
ridge, indicated by a dashed black line. It shows 
only a moderate slop in $\mu$, which explains why the maxima are more 
pronounced when we vary $\beta$ at fixed $\mu$ than vice versa.

The broadening of this ridge at small $\mu$ as well as of the
maximum in Fig. \ref{poleps} is responsible for the loss of a sharp
transition signal at small $\mu$. These figures clearly show that the
transition at fixed $\mu = 0.50$ is less steep than the one at
$\mu=0.80$. Presumably at $\mu < \sim 0.6$ we are dealing with a crossover,
whereas at large $\mu$ the signal is more compatible with
a real phase transition. Notice that
changing $\beta$ at fixed $\mu$, we cross the transition line at a more
oblique angle at smaller $\mu$, but the broadening of the ridge and
loss of a transition signal is a genuine effect, as can be seen from
fig. \ref{polland}.
A second ridge branching off from this main ridge at large $\mu$,
highlighted by a dotted line is suggested by looking at the level lines in
fig. \ref{polland} and corresponds to the second maximum suggested at
large $\mu$ in fig. \ref{poleps}. This may indicate the appearance of the 
new phase at large $\mu$ and small $T/T_c$ discussed above.

We used the results for the Polyakov loop susceptibility to estimate the
possible position of the transition points in the $\beta$ vs. $\mu$ plane
(see \cite{dfss} for details); to go half way toward a possible physical
interpretation the positions determined in this way are indicated by 
the blobs in the diagram 
$T/T_c$ vs. $\mu_{phys}/T_c$ of fig. \ref{polland}, where
$\mu_{phys}= \mu / a(\beta) = N_{\tau} \mu T$. 
The shaded blobs correspond
to the rather unambiguous `deconfining' signal observed for
$\mu > \sim 0.6$ ($\beta <\sim 5.72$). The `transition' line suggested
by this signal
 starts at the lower point A on the figure, located at
$\beta\simeq5.55,\,\mu\simeq 0.88$, i.e., with our rough estimation
$\mu_{phys}/T_c \simeq 2.4,\,T/T_c\simeq 0.45$ (below which we could no
longer obtain reliable data) and ends at the point B located near
 $\beta\simeq 5.72,\,\mu\simeq 0.6$, i.e., with our rough estimation
$\mu_{phys}/T_c \simeq 2.3,\,T/T_c\simeq 0.65$. Above this point the
signal becomes ambiguous.

The picture emerging from the data is thus the following: for $\mu < 0.5 -
0.6$ ($\mu_{phys}/T\sim 3$)  there is only a broad crossover, while for
$0.6 < \mu < 0.9$ ($3.6 < \mu_{phys}/T < 5.3$) there is evidence of a
sharper crossover or transition at a value $\mu_c$ depending on $\beta$.
Moreover, for $\mu \simeq 0.9$ there is some evidence of the presence of
the second transition even though this evidence is much weaker than the
other one because at larger values of $\mu$ the fermion determinant
strongly oscillates and, indeed, the usual sign problem manifest its
effects.

To summarize our results, 
the phase structure found by the numerical simulations for $n_f=3$
is shown in fig. \ref{polland}. The signal for the deconfining transition
(or narrow crossover) on the line connecting A and B is rather good and
it also appears that at small $\mu$ (above B) the transition is
smoothed out in accordance with the expectations from full QCD
simulations \cite{karrev,afks}. A second transition at large
$\mu$ could only be identified tentatively. In this region, the diquark
susceptibility grows strongly \cite{dfss}. This region needs further study to reach
a conclusion, but it is interesting that the general picture shows
qualitative agreement with the one found in the mean field approximation.

We can consider this model as an evolved `quenched
approximation' in the presence of charged matter. Then this study would
give us information about the modified gluon dynamics of the SU(3) theory
in this situation. It would then be natural to think of it as providing a
heavy, dense, charged background for propagation of light quarks and
calculate light hadron spectra and other hadronic properties under such
conditions. This could also help fixing a scale controlling the behavior
of the light matter. Work in progress goes in this direction.

\end{document}